\documentstyle[aps,prl,multicol,amssymb,epsf,float,psfig]{revtex}

\newcommand{\ket}[1]{\left | #1 \right \rangle}
\newcommand{\bra}[1]{\left \langle #1 \right |}

\title{Entanglement Simulations of Shor's Algorithm}
\author{S. Parker and M.B. Plenio}
\address{Optics Section, The Blackett Laboratory, Imperial College,
London SW7 2BW, England}
\date{\today}

\begin{document}
\draft
\maketitle
\begin{abstract}

We demonstrate that, in the case of Shor's algorithm for
factoring, highly mixed states will allow efficient quantum
computation, indeed factorization can be achieved efficiently with
just one initial pure qubit and a supply of initially maximally
mixed qubits (S. Parker and M. B. Plenio, Phys. Rev. Lett., {\bf
85}, 3049 (2000)). This leads us to ask how this affects the
entanglement in the algorithm. We thus investigate the behavior
of entanglement in Shor's algorithm for small numbers of qubits by
classical computer simulation of the quantum computer at different
stages of the algorithm. We find that entanglement is an intrinsic
part of the algorithm and that the entanglement through the
algorithm appears to be closely related to the amount of mixing.
Furthermore, if the computer is in a highly mixed state any
attempt to remove entanglement by further mixing of the algorithm
results in a significant decrease in its efficiency.

\end{abstract}


\begin{multicols}{2}

\section*{Introduction}

Quantum entanglement \cite{Plenio V98,Nielsen C01} is a basic resource in
quantum information processing. While its role in quantum communication tasks
is quite well understood the same cannot be said about quantum
 computation. While it is generally believed that entanglement is necessary to
achieve an exponential speedup of a quantum algorithm over a classical
algorithm \cite{Jozsa} the exact mechanism by which this may happen
 is unclear. In fact, so far it has not been proven strictly whether
entanglement is really necessary for an exponential speedup.

Generally, the
argument for the power of quantum computation relies on the assumption that any
algorithm that simulates the time evolution of a quantum system will be
exponential in the number of quantum bits involved, if it explores the whole
 state space. The reason is that the dimension of the total state space of a
 quantum system grows exponentially with the number of subsystems. For pure
states, this argument implies that the quantum system needs to evolve into an
 entangled state to be difficult to be simulated. If it is always in a product
state, then the number of parameters required to describe it grows only
polynomially with the number of subsystems. For mixed states, however, the
 situation changes significantly. The set of disentangled states, i.e.\@ the
separable states, has the same dimension as the set of all states, although its
relative size with respect to the total state space decreases rapidly
 \cite{TarrachVidal}. Therefore, one could imagine that there are dynamics that
always leave the system in a separable state, but which are nevertheless
difficult to simulate on a classical computer simply by the fact that the
number of parameters that is required to describe the quantum system grows
exponentially. As a consequence it may conceivably be possible to
have efficient quantum computations on separable states as the algorithm is able 
to efficiently simulate itself. Furthermore this points to the possibility that 
efficient quantum computation is possible on
mixed states.

Recently Knill and Laflamme \cite{Knill} investigated the power of quantum 
computations using one pure qubit and a supply of maximally mixed qubits and 
were able to construct a problem that could be solved more efficiently using 
these resources than any known classical algorithm. Also Schulman and Vazirani 
\cite{Vazirani} were able to show that given a supply of thermal states one 
could produce a single pure qubit together with many maximally mixed qubits. The 
latter could then be discarded and the pure qubit combined with other pure 
qubits in a quantum algorithm. Indeed NMR quantum computation \cite{NMR} does 
start with initially thermally mixed qubits, although the computation efficiency 
falls off exponentially with the number of qubits. However, we still have the 
possibility that these mixed qubits could be used in a useful computation like 
that of Knill and Laflamme. It is therefore of interest to
 explore this idea further and see what degree of mixing a quantum computer can
tolerate before it loses it's efficiency.

This paper is an exploration of this idea. We study the efficiency of Shor's 
algorithm when the quantum computer is in a highly mixed state. We arrive at the 
conclusion that Shor's algorithm can be run on extremely mixed states without 
significant loss of computational efficiency. Nevertheless, it turns out that 
despite the significant degree of mixedness Shor's algorithm runs through some 
weakly entangled states, leaving the question open as to whether a
quantum computer really requires entanglement to be efficient.

The sections of this paper are organized as follows: in section I we give an 
outline of Shor's algorithm together with possible gate layouts and 
interpretations; in section II we examine what is known about simulating quantum 
algorithms; section III looks at entanglement measures for mixed states and 
section IV multipartite entanglement and it's quantification; sections V and VI 
give details of the simulations used in this work and some of the results 
obtained and finally section VII gives concluding remarks.

\section{Outline of Shor's Algorithm}
\subsection{The Algorithm}
\label{salg}

Shor's algorithm for factoring an integer $N = pq$, where $p$ and $q$ are prime, 
relies on finding the period, $r$, of the function $f_a(x) = a^x \hbox{mod} N$, 
where $a$ is some integer less than and coprime to $N$ chosen at random. Then, 
with a sufficiently high probability, at least one of the unknown factors of $N$ 
is given by $\hbox{gcd} \left(a^{r/2} \pm 1, N \right)$ which can be calculated 
efficiently using Euclid's algorithm. Classically all known algorithms are 
unable to solve the period finding problem in time polynomial in $\log N$, the 
length of the number being factorized.

The quantum period finding algorithm works as follows: two quantum registers are 
required whose state spaces are of size at least $N^2$ and $N$ respectively. As 
we will be using qubits we will require $L = 2\lceil \log_2{N} \rceil$ and $n = 
\lceil \log_2{N} \rceil$ of these two level systems for the two registers 
respectively. We initially prepare the first register in an equal superposition 
of all possible states by preparing each of the qubits of the register in state 
$\frac{1}{\sqrt{2}} \left(\ket{0} + \ket{1}\right)$. The second register is 
prepared in state $\ket{1}$, where all the individual qubits are in state 
$\ket{0}$ except the first which is in state $\ket{1}$. We now unitarily 
transform the two registers with the transformation  $U\ket{x, b} \rightarrow 
\ket{x, b a^x \hbox{mod} N}$ so that
\begin{equation}
\label{U}U \frac{1}{\sqrt{t}} \sum_{x=0}^{t-1} \ket{x, 1} \rightarrow 
\frac{1}{\sqrt{t}} \sum_{x=0}^{t-1} \ket{x, a^x \hbox{mod} N}
\end{equation}
where $t = 2^{L}$. Now, an inverse quantum Fourier transform
\begin{equation}
\label{F-1}
F^{-1} \ket{y} \rightarrow \frac{1}{\sqrt{t}} \sum_{z=0}^{t-1} e^{-2\pi i y z/t} 
\ket{z}
\end{equation}
on the first register of Eq. \ref{U} yields the state
\begin{equation}
\label{final}
\frac{1}{t} \sum_{x=0}^{t-1} \sum_{z=0}^{t-1} e^{-2\pi i x z/t} \ket{z, a^x 
\hbox{mod} N}
\end{equation}
and the first register now contains information about the period of the function 
$f_a(x)$. We access this information by simply measuring the first register in 
the number, or {\em computational} basis obtaining the result $\ket{c}$, say. It 
was then shown in \cite{Shor95} that the fraction $c/t$ is, with a sufficiently 
high probability, most closely approximated (using the continued fractions 
method \cite{Summary}) by a fraction $j/k$ (with $k<N$) which in lowest terms 
has $k=r$, the period we are trying to find and will therefore sufficiently 
often give us a factor of $N$.

\subsection{Decomposition into basic gates}
\label{abcdecomp}

These are the essential details of Shor's algorithm as it was first formulated. 
We must now, of course, be sure that the algorithm runs in {\em time} polynomial 
in $\log{N}$ for general $N$ as well as using polynomial space as has been shown 
above. The time taken to perform the algorithm is generally assessed by counting 
the number of basic operations involved. To do this we must have a decomposition 
of the $U$ and $F^{-1}$ transformations into elementary gates acting on a small 
numbers of qubits, usually one, two or three.

One condition on these elementary gates is that each of them is reversible, that 
is, given the output states we could work out the input states (and obviously 
vice versa). The condition is imposed by the unitarity, and therefore the 
reversibility of the transformations $U$ and $F^{-1}$. Examples of such 
reversible gates are CNOT's (2 qubits), TOFFOLI's (3 qubits) and an infinite 
variety of 1-qubit gates and 2-qubit gates where a single qubit transformation 
is 'controlled' by a second qubit.

Detailed polynomially efficient gate layouts for the modulo exponentiation
transformations can be found in \cite{decomp}. They highlight one other
complication: the efficient decomposition of general unitary transformations
into small basic gates seems to require auxiliary qubits. These are used during 
the transformation to store quantum information temporarily but are left in 
their initial states at the end. Some of these must be prepared in a known state 
($\ket{0}$, say), others may be prepared in a completely unknown, or maximally 
mixed state which may or may not be entangled to other systems outside the 
computer. Either way they must be returned to their initial state after use 
during the computation. If they are not returned to their initial states (or 
some other known state that is disentangled from the rest of the computer) these 
qubits will be holding information about the states of the non-auxiliary qubits 
so that the transformation as viewed on the {\em non-auxiliary qubits alone} 
cannot be unitary or reversible and quantum information will have leaked out of 
the quantum computer.

A polynomially efficient gate decomposition of the (inverse) Fourier transform 
into 1-qubit Hadamard transformations and 2-qubit controlled phase rotations can 
be found in Fig. \ref{diag1} and is all the gates to the right of, and 
including, the first Hadamard transform (H). It requires $O\left( (\log{N}) 
^2\right)$ 1- and 2-qubit gates to perform the transformation and is therefore 
again polynomially efficient in time.

The careful (or experienced) reader will notice that for increasing $N$ the
conditioned phase rotations are of increasingly small values (controlled $R_L = 
\left( \begin{array}{cc} 1 & 0\\ 0 & e^{-2\pi i/2^L} \end{array} \right)$ 
transformations are used) which would require that the accuracy of the gate 
implementation is exponential in $\log{N}$. This would require exponential 
resources (in terms of time/energy etc.\@) but it clear that the Fourier 
transform implemented without performing the controlled phase shifts to such 
high accuracy does not affect the transformation too much and so does not reduce 
the efficiency too much. In return, however, in a practical situation involving 
decoherence it is an advantage not to carry out these small phase shifts as the 
computation will then suffer less errors due to it's shorter computation time 
\cite{BarencoBrun1996/1997}.

\subsection{The phase kickback interpretation}
\label{skickback}

In Fig.\@ \ref{diag1} the controlled modulo exponentiation of the classical 
number $a$ has been decomposed into $L$ successive controlled modulo 
multiplications by $a^{2^{L-1}}\, \hbox{mod} N, a^{2^{L-2}} \,\hbox{mod} N, 
\ldots, a^2 \,\hbox{mod} N, a \,\hbox{mod} N$ \cite{ek}. We will write these as 
${}_cU_a^{2^{L-1}}, {}_cU_a^{2^{L-2}}, \ldots, {}_cU_a^{2^1}, {}_cU_a$, where
\begin{eqnarray}
\label{us}
{}_cU_a^{2^x} \ket{0, b} &=& \ket{0, b} \nonumber \\
{}_cU_a^{2^x} \ket{1, b} &=& \ket{1, a^{2^x} b \, \hbox{mod} \, N}.
\end{eqnarray}
The modulo multiplications can be written in this way, as powers of the gate
$U_a$, because multiplication by $a^{2^x}$ modulo $N$ is equivalent to
multiplying by $a$ modulo $N$, $2^x$ times. Actually performing the modulo
multiplications in this way would of course require exponentially many
repetitions of the basic gate ${}_cU_a$ and is therefore a highly inefficient 
method but each of the controlled modulo multiplications can be performed in 
time polynomial in $\log{N}$ after classical  precalculation of the numbers 
$a^{2^x} \,\hbox{mod} N$ \cite{decomp}.
\end{multicols}
\begin{minipage}{6.54truein}
 \begin{figure}[H]
 \begin{center}
  \leavevmode
  \epsfxsize=6.20truein
  \epsfbox{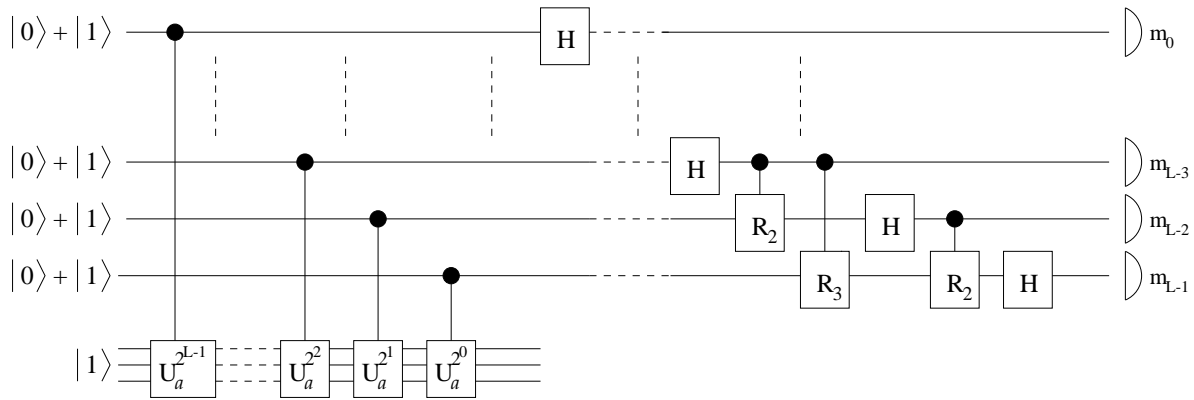}
 \end{center}
 \caption{An implementation of Shor's algorithm \protect\cite{ek}. The
controlled-$U_a$ operations produce controlled phase shifts related to the 
period of $f_a(x) = a^x \hbox{mod} N$ and the remaining Hadamard transformations 
(H) and controlled rotations $R_j = {1 \, 0 \choose 0 \, \phi_j }$ with $\phi_j 
= e^{-2 \pi i/2^j}$ implement the inverse Fourier transform. The result of the 
measurement, $c$, as described in section \ref{salg} is given by $c = 
\sum_{i=0}^{L-1} 2^i m_i$}
 \label{diag1}
 \end{figure}
\end{minipage}
\begin{multicols}{2}

Viewing the algorithm as such leads us to another interpretation of the period 
finding algorithm  that is just a change of basis away from Shor's original 
formulation, as outlined above. The operation $U_a$ has a set of $r$ eigenstates 
$\ket{\psi_j}$ ($j = 0,\ldots,r-1$) with eigenvalues $e^{2 \pi i j/r}$. Applying 
a controlled-$U_a$ gate to the state
\begin{equation}
\label{kick1}
\left(\ket{0} + \ket{1}\right)\ket{\psi_j}
\end{equation}
(aside form normalization) kicks the acquired phase onto the control qubit:
\begin{equation}
\label{kick2}
U_a \left(\ket{0} + \ket{1}\right)\ket{\psi_j} = \left(\ket{0} + e^{2 \pi i j/r} 
\ket{1}\right)\ket{\psi_j}.
\end{equation}
We cannot prepare the eigenstate $\ket{\psi_j}$ as this would require
knowledge of $r$. However, using the result $\sum_{k=0}^{r-1} \ket{\psi_j} =
\ket{1}$ \cite{ek} gives
\begin{equation}
\label{kick3}
{}_cU_a \left(\ket{0} + \ket{1}\right)\ket{1} = \sum_{k=0}^{r-1} \left(\ket{0} +
e^{2 \pi i j/r} \ket{1}\right)\ket{\psi_j}.
\end{equation}
A measurement of the control qubit, in some chosen basis, will now yield
information about the fraction $j/r$ for some $j$ selected at random, although 
only one bit of information will be acquired. More information can be acquired 
about the phase if we perform the controlled- $U_a^{2^{L-1}}, U_a^{2^{L-2}}, 
\ldots, U_a^{2^1}, U_a^{2^0}$ gates, using different control qubits for each, 
the inverse Fourier transform on the control qubits \cite{ek} and a projective 
measurement on each control qubit. This will sufficiently often allow us to 
obtain $r$ as described above: by finding the fraction $j/r$ closest to $c/t$ 
where c is the result of the measurements on the control qubits.

Shor's algorithm, then, can be seen in terms of the production and measurement 
of relative phase information which is related to $r$.

\subsection{Using mixed states}
\label{ssmix}

In \cite{ourpaper} it was shown that $U_a$ also has other sets of eigenstates 
$\ket{\psi^d_{j_d}}$ ($j_d = 0, \ldots, r_d-1$) with eigenvalues $e^{2 \pi i 
j_d/r_d}$ and that in fact nearly all of these (at least $(p-1)(q-1)$ of them) 
have $r_d = r$. Consequently the lower qubits (the 2nd register) in Fig. 
\ref{diag1} need not be prepared in the initial state $\sum_{k=0}^{r-1} 
\ket{\psi_j}= \ket{1}$ but can be prepared in the completely unknown state
\begin{equation}
\label{maxmix}
\frac{{\bf 1}}{N} = \frac{1}{N} \sum_{j_d, d} \ket{\psi_{j_d}^d}
\bra{\psi_{j_d}^d}
\end{equation}
(here we are equating $\ket{\psi_{j_1}^1} = \ket{\psi_j}$). This mixed state
algorithm is run exactly as before and the period is found at least
$\frac{(p-1)(q-1)}{N}$ times as efficiently as the original pure state
algorithm, this factor approaching unity as $p,q \rightarrow \infty$.

This also, in fact, means that any randomly selected state, whether pure or 
mixed, entangled or not, may be used as an input state for the lower qubits and, 
on average, the algorithm will run efficiently. These lower qubits may also be 
mixed because they are entangled to systems outside the computer 
\cite{ourpaper}.

In terms of the number of pure qubits that are needed in the algorithm we should 
once again address the matter of the decomposition of the  controlled-$U_a$ 
transformations into basic gates (Section \ref{abcdecomp}). It is well known 
that polynomially efficient decompositions of these transformations do exist 
\cite{decomp} but they require auxiliary qubits which it seems need to be 
prepared in a pure state. With some alterations to these decompositions, 
however, it has recently been shown that, of those auxiliary qubits that cannot 
be removed from the algorithm, only one need be prepared in a pure state 
\cite{thesis} (this is not to be confused with the `one' pure qubit of the 
Abstract and the next section). The rest can be prepared in maximally mixed 
states, although most can no longer be considered as being auxiliary to the 
computation.  

For simplicity, however, we will not consider these auxiliary qubits or this 
altered form of the controlled-$U_a$ transformations for the rest of this work.

\subsection{Using only one control bit}

One further modification to the algorithm is to use a {\em semi-classical Fourier 
transform} \cite{griff} - notice that the gates of the Fourier transform, both 
one and two qubit, occur sequentially on the qubits. We could thus replace all 
of the first register of qubits with a single control qubit and perform the gate 
operations as follows (see Fig. \ref{diag2}): we implement the first 
controlled-$U_a^x$ gate and the Hadamard transformation and measure the control 
qubit; after resetting the state of this qubit to $\ket{0} + \ket{1}$ we 
implement the next controlled-$U_a^x$ gate and replace the 2-qubit controlled 
phase shift with a {\em single} qubit phase shift {\em if the result of the 
first measurement was} $\ket{1}$; we continue in this manner with a Hadamard 
transformation, single qubit phase shifts given the results of {\em all previous 
measurements}, and another measurement and resetting. At the end of the 
algorithm we have a set of measurement results that have an identical
probability distribution to the algorithm using $L$ control qubits in the first 
register, as in Fig. \ref{diag1}.

\end{multicols}
\begin{minipage}{6.54truein}
 \begin{figure}[H]
 \begin{center}
  \leavevmode
  \epsfxsize=6.50truein
  \epsfbox{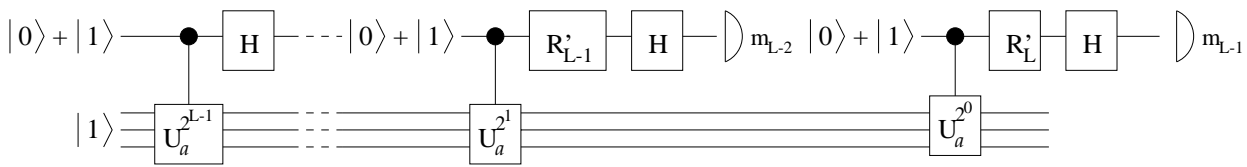}
 \end{center}
 \caption{An implementation of Shor's algorithm using only one control qubit
which is recycled. $R'_j$ are now combinations of the rotations $R_j$ in Fig.\@ 
\ref{diag1} given the results of previous measurements: $R'_j = {1 \, 0 \choose 
0 \, \phi'_j }$ with $\phi'_j = e^{-2 \pi i \sum^j_{k=2} m_{j-k}/2^k}$.}
 \label{diag2}
 \end{figure}
\end{minipage}
\begin{multicols}{2}

\section{Simulating algorithms}
\label{ssimalg}

We saw in the previous section how a quantum algorithm consisting of quantum
gates on quantum systems can factorize an integer $N$ in time and number of
qubits polynomial in $\log{N}$. The question then arises as to whether or not we 
can turn this into a classical algorithm by writing out the effect of the gates 
and measurements on  a classical system such as a computer or piece of paper. If 
we then find that we can do this efficiently (in time polynomial in the number 
of qubits) then we have an efficient classical algorithm for factoring. As no 
efficient algorithm is known we might be fairly certain that no such efficient 
simulation is possible and that all simulations of Shor's algorithm will be 
difficult.

Of course we already have a 'simulation' of Shor's algorithm as we have written 
down the equations (\ref{U}) and (\ref{final}) but we cannot in general derive 
any of the properties of the computer, or indeed the probability distribution of 
the measurement results, without writing out the density matrix for the whole 
system (or doing something else equivalently difficult) at the relevant point.

We would like to look at the system after certain gates or sets of gates. We can 
easily write out the effect of a single-qubit gate $A$ on a single qubit, 
whether pure or mixed, by writing down the unitary matrix for $A$ and working 
out its effect on the state vector $\bf v$ or density matrix $\rho$:
\begin{eqnarray}
\bf{v'} &=& A \bf{v} \label{matmul1} \\
\rho' &=& A \rho A^{\dagger}. \label{matmul2}
\end{eqnarray}
However, we can only do this if the qubit is completely disentangled from other 
systems which we may later wish to use for information processing.  This is 
because we clearly cannot, for example, simulate the operation of two single 
qubit gates on two qubits in an entangled state by tracing out the opposite 
qubit, implementing the gates on each and taking the combined state after the 
operations as the tensor product of the two states (this is not even the case if 
the two operations are the identity). That is, there are entries in the density 
matrix of the combined system which refer to the combined system rather than to 
the individual systems themselves.

Consider some examples in Shor's algorithm. The single-qubit gates in the 
Fourier transform are likely to be acting on qubits which are entangled to other 
qubits in the computer (possibly many of them) so to simulate the algorithm 
correctly we must not consider just the state of the single qubit (which will be 
mixed in general as it is entangled to other qubits). In contrast we noted in 
section \ref{ssmix} that we may use mixed states in the lower register of Shor's 
algorithm and that these may be mixed because they are entangled to systems {\em 
outside} the computer. In this case we {\em can} address only those qubits which 
are within the computer, even though they may be entangled to outside systems, 
because we will not later be concerned with these outside systems.

Let us consider the problem in more detail, considering pure states first. The 
state vector of an $M+1$-qubit system in a general entangled state requires $O 
\left( 2^{M+1} \right)$ complex numbers to be written down and stored 
\cite{Jozsa}. If we wish to classically simulate the application of the single-qubit gate 
$A$ to a qubit that is entangled to $M$ other qubits one method of simulation would be to 
apply the 2 by 2 matrix
representing the gate to each of the $M$ pairs of amplitudes in state vector corresponding to
different states of the remaining $M$ qubits. This involves $M$ 2 by 2 matrix multiplications 
and therefore requires $O \left( 2^M \right)$ operations



So to just simulate the effect of a single-qubit transformation in general takes 
classical resources exponential in $\log{N}$, if entanglement exists across the 
$O \left( \log{N} \right)$ qubits.

Of course this argument also follows for mixed states - we cannot simulate the 
effect of even one single-qubit gate efficiently if it is entangled to many 
other qubits within the computer. Using a method similar to that described above
we can perform the 2 by 2 matrix multiplication on $\left( 2^M \right)^2$ blocks
(pre-multiplication by $A$ and post-multiplication by $A^{\dagger}$) of the density
matrix for the whole computer, thereby requiring $O \left( 2^{2M} \right)$ operations.

But the situation here is more complicated: 
for pure states it is relatively easy to see if entanglement exists in a 
simulated algorithm (although it is by no means trivial) and whether therefore 
the above general method needs to be used to simulate gates acting just on parts of the
computer. For mixed states, however, it is harder to 
determine which qubits are separable. For two qubits to be separable there must 
{\em exist} a decomposition into pure states where the pure states are all 
separable:
\begin{eqnarray}
\label{sepcond}
\rho_{12} \,\, \hbox{separable} \,\, &\Leftrightarrow& \,\, \exists \,\,
\ket{\psi_i}_{12}, p_i (>0)\,\, \hbox{such that} \nonumber  \\
\rho_{12} &=& \sum_i p_i \left( \ket{\psi_i} \bra{\psi_i} \right)_{12} \,\,
\hbox{ where} \nonumber \\
\forall i \ket{\psi_i}_{12} &=& \ket{\psi^1_i}_1 \otimes \ket{\psi^2_i}_2
\end{eqnarray}

For almost all states there will be a decomposition into non-separable states. 
Finding a separable decomposition however, and indeed finding if it exists, is a 
difficult task.

A mixed state algorithm, then, may be seen as a mixture of pure state algorithms 
and even if at each stage these pure state algorithms are entangled there may 
exist other sets of pure state algorithms which are not entangled which mix to 
give the same mixed state algorithm. And these disentangled sets of pure 
algorithms will be different after each gate so we cannot easily preclude one 
existing \cite{Jozsaprivate}.

\section{Entanglement measures and mixed states}
\label{sentmeas}

\subsection{Entanglement measures for pure states}
\label{sseentmeaspure}

Let us first deal with how we would measure entanglement between two quantum
systems whose combined state is pure. Many entanglement measures in some way
view entanglement as a resource. The process of teleportation \cite{tele,tele1}
is an important example of a phenomenon that requires entanglement to be
observed at all, as it is required in many other applications of quantum
information processing including entanglement swapping, dense coding, precision
measurements and hiding classical information \cite{tele,othernonlocal} and in 
the
violation of Bell's inequalities \cite{Bellineq}.

To do perfect teleportation of an unknown single qubit state requires one of the
four Bell states (or 'EPR pairs') (Eq. \ref{Bellstate1} and \ref{Bellstate2}) to 
be shared between the two separated
parties 1 and 2:
\begin{eqnarray}
\ket{\phi^{\pm}_{12}} = \frac{1}{\sqrt{2}} \left( \ket{0}_1 \ket{0}_2 \pm
\ket{1}_1 \ket{1}_2 \right) \label{Bellstate1} \\
\ket{\psi^{\pm}_{12}} = \frac{1}{\sqrt{2}} \left( \ket{0}_1 \ket{1}_2 \pm
\ket{1}_1 \ket{0}_2 \right). \label{Bellstate2}
\end{eqnarray}
Any other state of two qubits (which cannot be transformed into one of the above 
by unitary transformations performed by the two parties separately) cannot be 
used to perform teleportation perfectly.

To quantify the entanglement of a general state $\ket{\psi}_{12}$
we could look at how well they perform the teleportation process (with some
sort of 'fidelity' measure between the input state and the output state or the 
maximum probability for perfect teleportation). In fact, we would rather ask  
how many Bell states
can be obtained from the given state $\ket{\psi}_{12}$ using only Local quantum
Operations (such as transformations, addition and removal of local separable
systems and measurements) and Classical Communication (LOCC for short)
\cite{procrust}. The Bell states, therefore, have entanglement of value '1'
because you can only obtain one Bell state from each Bell state supplied (you
cannot on average increase the number of Bell states with LOCC).

For other states of two qubits, say
\begin{equation}
\label{genpsi}
\ket{\psi}_{12} = a \ket{00}_{12} + b \ket{11}_{12}, \,\, a, b \in {\cal R}^+, 
\,\, a>b,
\end{equation}
(with $a^2 + b^2 = 1$) let us consider first what we can do with just one copy 
of the state. The most efficient method of obtaining a Bell state from this 
state is to use the Procrustean method \cite{procrust,finite1,finite2}. This 
only creates a Bell state with  probability $2|b|^2$, the rest of the time 
creating a completely separable state.

We can do better than this efficiency if  we are supplied with more copies, $n$ 
say, of the state $\ket{\psi}_{12}$ held by the separated parties. Now we can 
increase the number of Bell states obtained between the parties {\em per copy 
of} $\ket{\psi}_{12}$ {\em held} by allowing each of the separated parties to 
perform {\em joint} operations on those parts of the entangled states each 
holds. So say $k$ is the number of Bell pairs obtained, then on average, and 
with a suitable method, $k/n \ge 2b^2$. The exact results for any finite number
of copies are known \cite{finite2} and asymptotically, in the
limit of large n (provided we allow an arbitrarily small probability of error), 
the average number of Bell pairs obtained per copy for pure
states is given by
\begin{equation}
\label{entdef}
\frac{k}{n} = E\left(\rho_{12}\right) = S\left(\rho_1\right) = 
S\left(\rho_2\right)
\end{equation}
where $\rho_{12} = \left( \ket{\psi}\bra{\psi} \right)_{12}$, $S(\sigma)$ is the 
von Neumann entropy of the density matrix $\sigma$, and $\rho_1$ and $\rho_2$ 
are the partial density matrices of the first and second of the entangled 
particles. For the state of Eq.\@ (\ref{genpsi}) this gives $E \left( \left( 
\ket{\psi} \bra{\psi} \right)_{12} \right) = -a^2 \log{a^2} - b^2 \log{b^2} > 
2b^2$. These, and future results also hold for two party systems composed of 
individual
systems of more than two levels.

In the asymptotic limit (only)  the process is also true in
reverse: given $k$ Bell pairs and taking $k \rightarrow \infty$ we
can create from them, using LOCC, $k/E\left(\psi_{12}\right) = n$
copies of the state $\ket{\psi}_{12}$. We say, then, that for pure
states the asymptotic {\em entanglement of formation}, $E_F$ (the
number of Bell states we need to pay per copy of state
$\ket{\psi}_{12}$ we get in return) \cite{entform,form}, is the
same as the asymptotic {\em distillable entanglement} $E_D$ (the
number of Bell pairs we can distill out of the state
$\ket{\psi}_{12}$ per copy of $\ket{\psi}_{12}$ paid).

\subsection{Mixed states}
\label{ssmixentmeas}

For mixed states the situation is far less clear. $E_F$ and $E_D$
can be defined in the same way but for general mixed states they
are not equal - you cannot in general get as many Bell pairs out of a state as
you would put in (you certainly cannot obtain more or you would be able to 
locally create
 entanglement {\em ad infinitum}). This is due to the fact 
that a mixed state is to some
extent unknown and the randomness inserted on the formation of the
state from Bell states cannot be eliminated unless extra
information about the state is obtained.

Another problem with $E_F$ and $E_D$ defined in this way is that
at present there are no analytical methods for calculating either
for general mixed states. The only example where an analytical
expression does exist for more than some specific subclasses of
states is the entanglement of formation of a single two-qubit
system \cite{form}.

These entanglement measures are not the only possibilities we
could produce. Others exist such as the relative entropy
\cite{relent}. So what are the conditions for a mathematical
object to be called an entanglement measure? One set of conditions
that are generally accepted as being sensible for an entanglement
measure of a state $\rho_{12}$ are as follows \cite{entcond}:

(i) $E\left(\rho_{12}\right) = 0$ if $\rho_{12}$ is separable.

(ii) Local unitary transformations on $\rho_{12}$ leave
$E\left(\rho_{12}\right)$ invariant, i.e.\@ $E\left(\rho_{12} \right) = E\left(
U_1 \otimes U_2 \rho_{12} U_1^{\dagger} \otimes U_2^{\dagger} \right)$.

(iii) $E\left(\rho_{12}\right)$ cannot, on average, increase under local 
operations and
classical communication.

We may also wish to add one more condition to this list. This says that

(iv) the entanglement measure should be equal to the pure state entanglement
measure (Eq. (\ref{entdef})) for all pure states.

Of course not all entanglement measures need obey this condition,
indeed the one we will be using below does not. There are many
candidates for entanglement measures and all these measures will
not agree with each other for mixed states. More importantly, the
measures will put a different {\em order} on the states
\cite{Shash,Jens}, that is, according to one measure of
entanglement $E_1$ the state $\rho_{12}$  may be more entangled
that $\sigma_{12}$ but according to another measure of
entanglement $E_2$  the state $\sigma_{12}$ could be more
entangled than $\rho_{12}$:
\begin{eqnarray}
\label{entorder1}
E_1 \left( \rho_{12} \right) > E_1 \left( \sigma_{12} \right) \nonumber \\
E_2 \left( \sigma_{12} \right) > E_2 \left( \rho_{12} \right). \nonumber
\end{eqnarray}
Indeed, it was shown in \cite{Shash} that any two entanglement
measures that agree for pure states but not for mixed states {\em
must}  put a different order on the states.

We must accept, then, that entanglement measures may put different
orders on states. The only alternative is to declare one
entanglement measure as the 'correct' one, which immediately
prevents us from examining how we might prepare entanglement and
how we might use it.

\subsection{A measure of entanglement for mixed states}
\label{ssmeasmix}

For the present work we need a measure that is easy to calculate
and has an analytical form for general mixed states. We will call
this measure the {\em logarithmic negativity} $E_{neg}$
\cite{Vidal,Werner,newmartin}. It is defined as follows: first
denote the matrix elements of the density matrix $\rho_{12}$ {\em
in some tensor product basis} by
\begin{equation}
\label{matele}
\rho_{12}^{ij,kl} = {}_1\bra{i} {}_2\bra{j} \rho_{12} \ket{k}_1 \ket{l}_2.
\end{equation}
The {\em partial transpose} \cite{NPT} with respect to system 2 is then defined
in this notation as
\begin{equation}
\label{PT}
\left( \rho_{12}^{T_2} \right)^{ij,kl} = \rho_{12}^{il,kj}
\end{equation}
(the labels $l$ and $j$ have swapped places). It has been shown
\cite{NPT} that the positivity of this new matrix (that is, the
positivity of all its eigenvalues) is a necessary condition for
the state to be separable. Therefore, $E_{neg}$ is now defined as
the log of the sum of the absolute values of the eigenvalues of
the new matrix $\rho_{12}^{T_2}$ or $\rho_{12}^{T_1}$ (the eigenvalues and therefore this
measure are also independent of the particular tensor product basis in which the state is
considered). This can be written in more compact form as
\begin{equation}
\label{Enegtrnorm}
E_{neg} = \log {Tr |\rho_{12}^{T_2}|} = \log {Tr |\rho_{12}^{T_1}|}.
\end{equation}
As mentioned above this measure does not agree with the pure state entanglement 
measure of Eq. (\ref{entdef}). However, one particularly useful property of 
$E_{neg}$ is that it is an upper bound for $E_D$:
\begin{equation}
\label{Ebound}
E_{neg} \ge E_D.
\end{equation}
What is more important is that if $E_{neg} = 0$ we can be sure
that the two party system does not have distillable entanglement,
although it is not known whether the reverse statement is true or
not: we cannot say that distillable entanglement does exist if
$E_{neg} \ne 0$. Also the fact that $E_{neg} = 0$ does not mean
that the state is separable (there exist states with $E_{neg} = 0$
that are inseparable, which are known as {\em bound entangled
states} \cite{bound} as the entanglement cannot be distilled into
Bell states but is somehow bound from us).

\section{Multipartite entanglement}
\label{smulti}

Entanglement does not only exist between two-party (bipartite)
systems, it can also exist between three or more parties. One
example is the three-party GHZ state \cite{GHZ}:
\begin{equation}
\label{GHZdef}
\ket{\psi_{GHZ}}_{123} = \frac{1}{\sqrt{2}}\left( \ket{0}_1 \ket{0}_2 \ket{0}_3
+ \ket{1}_1 \ket{1}_2 \ket{1}_3 \right).
\end{equation}
One particular property of this state is that it has no bipartite
entanglement in the sense that if we trace out the third party, say, the
remaining bipartite state is given by
\begin{eqnarray}
\label{twoparty}
Tr_3 \left( \ket{\psi_{GHZ}} \bra{\psi_{GHZ}} \right)_{123} \nonumber \\
= \frac{1}{2} \left( \ket{0}_1 \bra{0} \otimes \ket{0}_2 \bra{0} + \ket{1}_1
\bra{1} \otimes \ket{1}_2 \bra{1} \right).
\end{eqnarray}
The entanglement in this state is of different nature to the one contained in
an EPR state because it is impossible to inter-convert GHZ states and EPR pairs
reversibly \cite{Popescu}. While one can create a GHZ state from two EPR pairs,
one can only ever obtain one EPR pair from one GHZ state. What is not known
for three party systems is whether GHZ states and EPR pairs are the only different kinds of
entanglement. What is known is that there are in fact more types of
multipartite entanglement for systems of {\em more} than three parties 
\cite{Galvao,Dur,Zhang}. Three-party entanglement, then, may contain two-party 
entanglement and under certain conditions we can locally reversibly (under
LOCC) transform between the two \cite{multi}.

How then do we measure the entanglement of a multipartite system? Because of the 
different types of entanglement involved having one general measure for all 
these types is difficult (the relative entropy suitably redefined for 
multipartite systems is perhaps one exception \cite{relentmultipart} although 
there are still many problems).

Our approach will be to use a bipartite entanglement measure to verify the 
existence of entanglement between all the different qubits in the computer. Let 
us suppose that we have an entanglement measure for bipartite systems, or a way 
of verifying whether a state is separable or not between the two parts of the 
bipartite system. We now use this measure on all possible bipartite 
partitionings of an $n-party$ system. For example, for a system of 4 qubits 
labeled as 1, 2, 3, and 4 the possible bipartite partitionings are
\begin{eqnarray}
\label{splits}
&&1 / 2 3 4, \,\, 2/134, \,\, 3/124, \,\, 4/123, \,\, \nonumber \\
&&12/34, \,\, 13/24, \,\, 14/23 \nonumber
\end{eqnarray}
where $1/234$ means system 1 is considered as being partitioned from systems 2, 
3 and 4. For general $n$ there are $2^{n-1} - 1$ such partitionings.

Clearly, if there is entanglement between the two sides for at
least one partitioning, then the state is entangled. The question remains as to 
whether the converse is true, that is, if all bipartite partitionings are 
separable is the state completely separable i.e.\@ the multipartite state can be 
written as a mixture of product states? For pure states it is clear that this is 
also true (in fact, you need only look at some particular subset of the possible 
partitionings). But for mixed states it is {\em not} true - there are states 
that are separable across all bipartite partitionings (and therefore have 
positive partial transpose (PPT)) but that are not completely separable 
\cite{uncomplete}. However this entanglement cannot be distillable entanglement 
- if we could distill it into pure multipartite entanglement by local operations 
it could in turn be changed into bipartite entanglement between qubits.

\begin{figure}
\begin{center}
\leavevmode
\centerline{\hbox{\psfig{figure=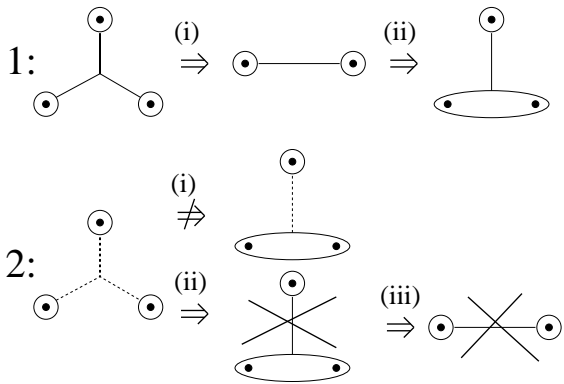,height=5.cm}}}
\end{center}
\caption{Diagram outlining the possible entanglements for a multi-particle 
system, here for three parties. (1.)\@ A three-party state which has distillable 
entanglement (solid line) (1.i.) can always be transformed into two-party 
distillable entanglement (at least some of the time). (1.ii.) This in turn means 
that distillable entanglement must exist across a bipartite boundary (one in 
some way separating the two parties who can obtain the two-party entanglement 
above) simply because allowing the third party to combine his operations 
non-locally with one of the other two is a more powerful operation. However, 
(2.) three-party non-distillable entanglement (dashed line) (2.i.) may or may 
not contain non-distillable entanglement between a two-party partitioning, 
although (2.ii.) it certainly does not contain distillable entanglement of this 
form and therefore (2.iii.) no distillable entanglement of any sort exist 
between two parties.}
\label{entfig}
\end{figure}

Our method in the simulations, then, will be to use the {\em logarithmic 
negativity} measure across all bipartite partitionings of the qubits in the 
algorithm. This will tell us if any distillable bipartite entanglement exists in 
the algorithm but will also show us if distillable multipartite entanglement of 
any form exists. This follows from the fact that this measure in effect verifies 
whether the state is PPT and is therefore not distillable across any bipartite 
partitioning. If this is the case then no multipartite distillable entanglement 
can exist (we cannot distill the entanglement into pure state entanglement) 
otherwise we would again be able to distill this into (pure) bipartite 
entanglement of some form.

So, we can indeed verify in this way whether any distillable entanglement of any 
form exists although we cannot preclude that there exist non-distillable 
entangled states.

\section{The Simulations}

\subsection{The Basic simulations}
\label{ssbasics}

Let us first introduce the basic method of our simulations. The states and gates 
are all stored in matrix form, the former as density matrices (because we will 
in general be using mixed states) and the gates as specific unitary 
transformations. $n$ qubits therefore require $2^n \times 2^n$ density matrices 
and unitary transformations. After inputting the initial state of the computer we 
simulate the effect of each gate by pre- and post-multiplying the density matrix 
by the unitary transformation and its Hermitian conjugate representing the gate 
to obtain the new state of the computer.

\subsubsection{Simulation of quantum gates}
\label{ssssimqgates}

We will not simulate the effect of all single-, two- or three-qubit gates but 
will only simulate the algorithm as it appears in Figure \ref{diag2} where we
have convenient points at which to examine the computer i.e.\@ we 
will only simulate the controlled modulo multiplication gates as a single gate 
and the gates of the Fourier transform. We will also be using this version of the 
algorithm as it reduces the number of qubits needed (by about 2/3) which will 
result in a great decrease in time and space resources the algorithm requires to 
be simulated.

It was noted in section \ref{ssimalg} that, in general, because of the potential 
entanglement across all the qubits in the algorithm, the simulation of even a single 
qubit gate requires an exponential number of operations.  For example, 
if the single qubit Hadamard transform
\begin{equation}
\label{H}
H = \frac{1}{\sqrt{2}}\left( \begin{array}{cc} 1 & 1\\ 1 & -1 \end{array} 
\right)
\end{equation}
is acting on the first qubit in a computer consisting of 2 qubits then the 
unitary transformation required is
\begin{equation}
\label{Hcross1}
H \otimes {\bf 1} = \frac{1}{\sqrt{2}}\left( \begin{array}{cccc} 1 & 1 & 0 & 0 
\\ 1 & -1 & 0 & 0 \\ 0 & 0 & 1 & 1 \\ 0 & 0 & 1 & -1\end{array} \right),
\end{equation}
and if it is acting on the second qubit
\begin{equation}
\label{1crossH}
{\bf 1} \otimes H =
\frac{1}{\sqrt{2}}\left( \begin{array}{cccc} 1 & 0 & 1 & 0 \\ 0 & 1 & 0 & 1 \\ 1 
& 0 & -1 & 0 \\ 0 & 1 & 0 & -1\end{array} \right).
\end{equation}

To simulate the effect of these gates on the 2 qubit state by straight matrix multiplication
however is not optimal. From the form of the above transformations it is clear that it is 
better to act with the Hadamard transform on blocks. In the case of
$H$ acting on the first qubit the density matrix written in block form evolves as
\begin{equation}
\label{toad}
\rho = \left( \begin{array}{cc} A & B \\ C & D \end{array} \right) \rightarrow 
\left( \begin{array}{cc} HAH^{\dagger} & HBH^{\dagger} \\ HCH^{\dagger} & HDH^{\dagger} 
\end{array} \right).
\end{equation}
Here of course $C = B^{\dagger}$ (saving us some calculation time) as it is a density matrix
and also for the Hadamard
transformation $H = H^{\dagger}$. The same is true for $H$ acting on the second qubit except
the elements in the 2 by 2 blocks upon which $H$ acts (within the 4 by 4 density matrix) 
are separated as is shown diagrammatically here:
\begin{equation}
\label{toad2}
\left( \begin{array}{cccc} a & b & a & b \\ c & d & c & d \\ a & b & a & b \\ c & d
& c & d \end{array} \right)  
\end{equation}
(lower case letters correspond to elements of blocks).

For a general $2^n$ by $2^n$ density matrix the simulation of a single qubit gates thus 
requires $O \left( 2^{2n} \right)$ operations. For gates acting on $m \leq n$ qubits this
number rises to $O \left( 2^{2n+m} \right)$ operations.

\subsubsection{Simulation of tracing and measurements}
\label{ssssimmeas}

First of all we may need to trace out one (or more) of the qubits (labeled by $x$) of 
the computer to leave the density matrix
\begin{equation}
\label{meastrace}
\rho_{12 \cdots (x-1)(x+1) \cdots n} = Tr_{x} \left(  \rho_{12 \cdots n} 
\right).
\end{equation}
In matrix form this tracing step effectively involves adding two $2^{n-1}$ by $2^{n-1}$
sub-matrices together  and zeroing the rest of the $2^n$ by $2^n$ matrix. 
Thus the tracing step requires $O \left( \left( 2^{n} \right) 
^2 \right)$ operations (although these operations are additions rather than multiplications and
are therefore considerably quicker). 

We can simulate single-qubit projective measurements in a similar way to that of 
simulating gates. We will assume, without loss of generality, that all measurements are done in the 
computational basis, as to perform a measurement in a different basis (even an 
entangled basis) we can unitarily transform the system (with entangling gates if 
necessary) and measure in the computational basis. 


First of all we must calculate the probabilities of a measurement on the qubit yielding the
result $\ket{0}$ ($p_0$) or $\ket{1}$ ($p_1 = 1-p_0$). This is done by 
summing those elements on the diagonal of the density matrix for the whole computer which
correspond to the measurement result.

For a Monte Carlo simulation of measurement results we may now generate a random number, 
$p$,  from a uniform linear distribution between 0 and 1, and if $p<p_0$ take 
the simulated measurement result to be $\ket{0}$, otherwise it is $\ket{1}$.

The measurement changes the state of the computer. To simulate this we must use 
two projection operators
\begin{equation}
\label{proj}
P_0 = \left( \begin{array}{cc} 1 & 0\\ 0 & 0 \end{array} \right) \,\, \hbox{and} 
\,\, P_1 = \left( \begin{array}{cc} 0 & 0\\ 0 & 1 \end{array} \right)
\end{equation}
and act on the state with either $P_0$ (if the measurement result was $\ket{0}$) or $P_1$ (if
the result was $\ket{1}$) just as we would act with a single qubit gate. This will, of course,
result in just setting 3/4 of the element of the density matrix to zero. This gives us the 
(subnormalised) state of the measurement collapsed computer, including the 
(partial or total) collapse of any qubits that are entangled to the measured qubit. 

We must then be sure to renormalize the state of the whole
computer. This can be done easily by dividing each entry of the
density matrix for the collapsed computer by the trace of the
density matrix (or equivalently the measurement probabilities, $p_0$ or $p_1$).

The simulation of the measurement therefore takes $O \left( 2^{2n} \right)$ operations but 
notice that the quantum computer does this in linear time, as it only needs to find 
qubit $x$ and measure it.

For the full Monte-Carlo type simulation we must of course repeat
the simulation of the algorithm a number of times and average any
properties of the system we obtain during each simulation.

\subsubsection{Other processes}
\label{sssotherproc}

We can also re-prepare the measured qubit in the required state
using this method by another application of single qubit gates. If
we wish to prepare the state $\ket{0} + \ket{1}$ then if the
measurement result was $\ket{0}$ we apply $H$ and if it was
$\ket{1}$ we apply a state flip $\left( \begin{array}{cc} 0 & 1\\
1 & 0 \end{array} \right)$ followed by $H$.

We can now simulate any algorithm we wish with any set of gates
and  any type of measurement (a Positive Operator Valued Measure
(POVM) can be simulated by adding auxiliary systems, performing
unitary transformations on these systems together with the
computer and measuring the auxiliary systems).

\subsubsection{Entanglement calculations}
\label{sssentcalc}

We are mainly interested in the degree on entanglement at each stage. As 
mentioned in section \ref{ssmixentmeas} the degree of entanglement does not 
change under local unitary operations so we need only examine the entanglement 
after operations on more than one qubit. From Fig. \ref{diag2} this will be 
after the controlled-$U_a$ operations (where entanglement may increase of 
decrease) as well after the measurements (where the entanglement may also 
increase or decrease but {\em on average} it should never increase - local 
projective operations can never increase the amount of entanglement on average).

As in section \ref{smulti}  we will use the logarithmic negativity measure of 
section \ref{ssmeasmix}, calculated and averaged across all possible bipartite partitionings 
of the system. If we are using a Monte Carlo simulation these must of course be 
averaged across repeated simulations of the algorithm as different entanglements 
will be observed with different measurement results.

\subsubsection{Efficiency Accountancy}
\label{ssseff}

Let us now check the time efficiency (or lack of it) of Shor's algorithm 
simulated by the above method:

(i) to form those gates acting on all qubits requires $O  \left(  2^{2n} 
 \right)$ steps and there are $O(n)$ of these giving $O \left( n 2^{2n} \right)$ steps altogether

(ii) for these gates the matrix multiplications for the simulation of the gate
require $O \left( 2^{3n} \right)$ operations. There are $O \left( n \right)$ of these gates
giving $O \left( n 2^{3n} \right)$ operations in all.

(iii) each application of a single qubit gate (including 
measurement projections) takes $O  \left( 
2^{2n} \right)$ steps and there are $O(n)$ of these, which is 
$O  \left( n 2^{2n} \right)$ operations altogether.

(iv) at $O(n)$ of the steps we wish to examine the entanglement $O \left( 2^{n} 
\right)$ times. This requires $O \left(  2^{2n} \right)$ steps 
each for the partial transposition, $O \left(  2^{3n}  \right)$ 
for the numerical eigenvalue routine \cite{numrec} and $O(n)$ steps for the 
remaining calculation of our entanglement measure. From the most inefficient 
process (numerical eigenvalue calculation) this gives $O \left(n  2^{4n} 
\right)$ operations for the whole. This then is the most important 
stage of the simulation in terms of the efficiency of the algorithm.

\subsection{Tree simulations}
\label{sstree}

In the above accounting we have not included any contribution from the fact that 
we have to repeat the whole simulation many times for the Monte Carlo method 
(section \ref{ssssimmeas}). Let us examine this more carefully. If we wish to 
estimate the probability distribution $\{P,1-P\}$ with two possibilities to a 
certain accuracy, $\epsilon$, we must repeat the Monte Carlo simulation 
approximately some number $T(\epsilon)$ times. At the $j^{th}$ measurement of 
the algorithm there would be $2^{j-1}$ states the computer could be in at the 
point (from the previous possible measurement results). For each of these we 
would need to estimate the probability distribution giving $2^{j-1} \times 
T(\epsilon)$ simulations we need to perform, this number getting exponentially 
worse at each step. So, given a fixed number of simulations the accuracy of the 
probability estimate and therefore the estimate of any properties of the system 
can get exponentially worse at each measurement step.

This leads us to an alternative simulation method, a 'tree' type simulation 
where every possibility of each measurement is considered. We have $L$ 
measurements during the algorithm and at each one the algorithm is given two 
possible 'paths' to use {\em for each of the paths already available}. So the 
computer can take two different paths (be in two different  states) at the first 
measurement, four different paths at the second and $2^L$ ($L = 2n$) different 
paths after the last measurement. This gives us $2^{2n+1} - 1$ possible states 
of the computer through the algorithm. And we want to sample the entanglement in 
the computer at two stages between each measurement, after the measurement and 
after the controlled modulo multiplication gates.

\begin{figure}
\begin{center}
\leavevmode
\centerline{\hbox{\psfig{figure=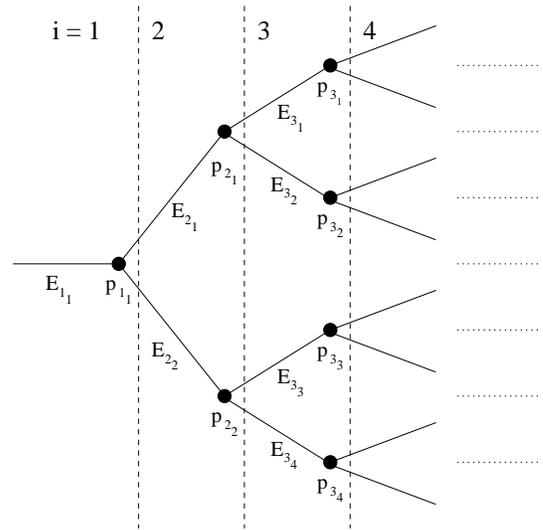,height=7.cm}}}
\end{center}
\caption{The 'tree' method of simulating an algorithm. $p_{i_j}$ are the 
probabilities for obtaining a measurement result $\ket{0}$ for the $j^{th}$ 
branch before the $i^{th}$ measurement ($j = 1 \ldots 2^{i-1}$). Likewise, 
$E_{i_j}$  are the values of some property, $E$, of the system for the $j^{th}$ 
branch before the $i^{th}$ measurement. The average of the property before the 
$i^{th}$ measurement is therefore $E^{av}_{i} = \sum_{j = 1}^{2^{i-1}}  \left( 
\prod_{k=1}^{i-1} p_{k_{\lceil j/2^{i-k} \rceil}} E_{i_j} \right)$.}
\label{treediag}
\end{figure}

We can calculate the probability of each of the measurement results at each 
stage of the algorithm and find the probability for each possible path leading 
to that stage. Having calculated these we can calculate exact averages for any 
properties of the system at any stage if we know the properties we wish to 
examine at each branch of the tree.

This gives us, of course, another exponential overhead in our simulation but, as 
we saw, this was the same for the Monte Carlo simulation.

\subsection{Noise simulation}
\label{ssnoise}

The high susceptibility of quantum computers to noise from the environment
\cite{noise} in which the computer is run has been known for some time.
This noise comes from entanglement of the computer with the
environment suffered during the execution of the algorithm. Further sources of
error are inaccuracies in the measurements and implementation of the gates.
Fortunately it has been shown that {\em error-correcting codes} exist
\cite{error}. These encode the state of a qubit into the joint (entangled)
state of many qubits such that random errors occurring  independently on the
qubits below a certain (reasonable) threshold can be corrected back to the
correct state using measurements and unitary transformations based on the
measurement results. It has also been shown that these codes can be used
in {\em fault tolerant} quantum error correcting schemes \cite{faulttol}, that
is, the measurements and transformations that implement the error correcting
code itself need not be implemented perfectly.

The fact that Shor's algorithm can be implemented with 'noisy' mixed states 
leads us to asking if this in any way increases the algorithm's robustness to 
noise, or whether the coherence within the pure state decomposition of the mixed 
states needs to be preserved.

It will be interesting, then, to simulate the effects of random noise injected 
into the computer and it's effect on the efficiency with which the number $N=pq$ 
is factorized.

There are many ways of simulating noise and many types of noise we could use in 
the simulation. We have selected two types, noise by measurement and noise by 
random Pauli operations. It should be noted, however, that different types of 
noise, whether in quantum or classical scenarios, tend to have similar 
qualitative effects and so we need not worry too much about the precise nature 
of each.

\subsubsection{Noise by measurement}
\label{sssnoisebymeas}

Here, we address each qubit in turn and with a given probability apply a 
measurement on that qubit in the computational basis (although it need not be in 
this basis for general noise). The probability for the two outcomes is governed 
by the state of the particle, as described in section \ref{ssssimmeas}.

This collapses the state of the computer in some way. We will, with the given 
probability, apply the noise step to every qubit after each gate during the 
algorithm. Of course, the controlled modulo multiplication gate consists of many 
gates acting on small numbers of qubits so would have more time to be effected 
by noise.

We should also carefully note that when running the algorithm we would not know 
the result of the measurement taken by the measurement noise and the algorithm 
would at that stage become more mixed (the computer becomes a weighted mixture of the two
states post measurement) as we do not know which measurement result
was found. In fact we would not even know if a noise step had been applied.

\subsubsection{Noise by Pauli operations}

Here, again addressing each qubit in turn, with a certain probability we will 
apply one of the three Pauli operators ($\sigma_x, \sigma_y, \sigma_z$), or the 
identity operator ($I$):
\begin{eqnarray}
\label{pauliops}
\sigma_x = \left( \begin{array}{cc} 0 & 1\\ 1 & 0 \end{array} \right)
&&\sigma_y = \left( \begin{array}{cc} 0 & -i\\ i & 0 \end{array} \right) 
\nonumber \\
\sigma_z = \left( \begin{array}{cc} 1 & 0\\ 0 & -1 \end{array} \right)
&&I        = \left( \begin{array}{cc} 1 & 0\\ 0 & 1 \end{array} \right)
\end{eqnarray}
to the qubit, these four operations occurring with equal probability.

Again we should be careful to point out that a noise step would leave the
computer in a equal mixture of the four states that result after the application of
the four Pauli operations. Thus the qubit will have its 
state completely randomized (although it could still be entangled to other 
qubits).

\subsection{Mixing of the control qubit}
\label{ssmixing}

As we want to examine the entanglement in the computer it will be interesting to 
investigate ways of reducing the entanglement and seeing how this affects the 
computer. We have already introduced mixed states into part of the quantum 
computer and seen that it does not affect the efficiency very much. But we could 
also mix the state of the control qubit. This must affect the efficiency of the 
algorithm (a totally random algorithm cannot be any use to us) and it will be 
useful to compare this to the change in entanglement.

We will do this by preparing (and re-preparing) the control qubit in the state
\begin{equation}
\label{contmix}
(1-\epsilon) \ket{\psi} \bra{\psi} + \epsilon \ket{\psi^{\perp}} 
\bra{\psi^{\perp}}
\end{equation}
where $\ket{\psi} = \frac{1}{\sqrt{2}} \left(\ket{0} + \ket{1}\right)$ and 
$\ket{\psi^{\perp}} = \frac{1}{\sqrt{2}} \left(\ket{0} - \ket{1}\right)$. 
Using the methods of our quantum computer simulator this tensor product 
operation is most conveniently done as follows: we assume that the 
control qubit is in state $\ket{0}$ (if it is in the state $\ket{1}$ after 
measurement it can be flipped easily); if we denote the state of the whole 
computer (including the control qubit) by $\rho_{12\cdots n}$ we calculate the 
temporary density matrix
\begin{equation}
\label{rhotemp}
\rho_{12\cdots n}^{\hbox{t}} = \left(F \otimes {\bf 1}^{n-1} \right)\, 
\rho_{12\cdots n} \, \left( F \otimes {\bf 1}^{n-1} \right);
\end{equation}
where $F$ denotes the single qubit flip operator (section \ref{sssotherproc}); 
we can then mix this with the original density matrix for the computer in the 
required proportions:
\begin{equation}
\label{mixit}
\rho_{12\cdots n}^{mix} = (1-\epsilon) \rho_{12\cdots n} + \epsilon 
\rho_{12\cdots n}^{\hbox{t}}
\end{equation}
and a final Hadamard transformation on the control qubit gives us the required 
state of the computer.

Changing the parameter $\epsilon$ between 0 and $1/2$ allows us to decrease or 
increase the mixedness of the computer.

\section{Results}
\label{sresults}

\subsection{Entanglement in Shor's algorithm}
\label{ssentres}

\begin{figure}
\begin{center}
\leavevmode
\centerline{\hbox{\psfig{figure=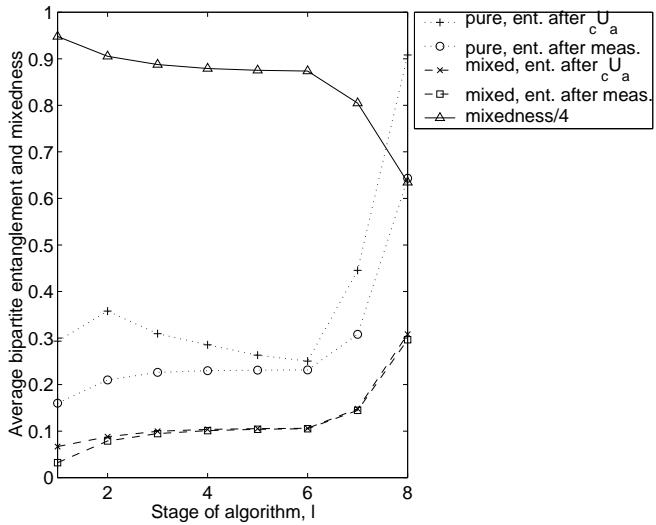,height=7.cm}}}
\end{center}
\caption{The average bipartite entanglement measured by the {\em logarithmic 
negativity} measure vs.\@ the stage of the algorithm for both pure and mixed 
state algorithms. The average entanglement is given after $s^{th}$ controlled 
modulo multiplication gate and after the $s^{th}$ measurement. This average is 
an average entanglement over all bipartite partitionings of the $n$-qubit system 
as well as over all algorithms for factorizing numbers which are products of two 
primes and have four binary digits (i.e.\@ 9, 10, 14 and 15) and over all 
possible numbers $a$ coprime to each of these numbers. Also shown is the 
mixedness (divided by 4) after the $s^{th}$ measurement. Notice how this closely 
mirrors the entanglement in the mixed state algorithm. }
\label{entn5}
\end{figure}

\begin{figure}
\begin{center}
\leavevmode
\centerline{\hbox{\psfig{figure=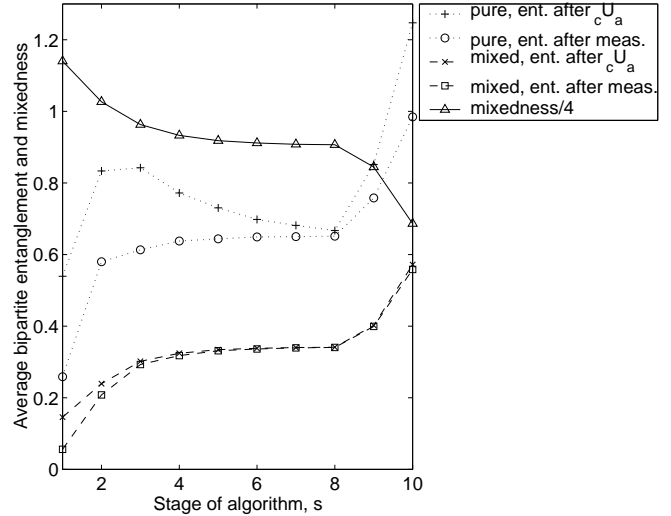,height=7.cm}}}
\end{center}
\caption{As Fig. \ref{entn5} but averaged over all algorithms for factorizing 
numbers which are products of two primes and have five binary digits (i.e.\@ 21, 
22, 25, 26) and over all possible numbers $a$ coprime to each of these numbers.}
\label{entn6}
\end{figure}

Firstly we will look at the entanglement in the pure and mixed state single 
control qubit algorithms (calculated using the 'tree' simulation method (section 
\ref{sstree})). We will look at the average bipartite entanglement as measured 
by the logarithmic negativity entanglement measure (section \ref{ssmeasmix}) 
averaged across all possible bipartite boundaries at each stage of the algorithm 
(section \ref{smulti}). We will also average across all possible algorithms 
factoring numbers of 4 binary digits and 5 binary digits. These results are 
shown in Figs. \ref{entn5} and \ref{entn6}.

These two sets of results have a very similar form, in particular the 
entanglement in the mixed state algorithms closely mirrors the mixedness of the 
quantum computer (the von Neumann entropy of the state of the whole computer) at 
each stage. Also note that the amount of entanglement increases towards then end 
of the algorithm where the most significant (i.e.\@ the highest) bits of the 
number $c$ are decided and the least significant bits of the period $r$ are 
found.

We can see immediately that in all algorithms, both pure and mixed, according to 
our entanglement measure, entanglement exists although it is up to three times 
lower in Fig. \ref{entn5} and \ref{entn6} for the mixed state algorithm. The 
mixed state algorithms are, however, at least around half as efficient as the 
pure state algorithms for the values of $N$ considered (which can be seen by 
calculating $\frac{(p-1)(q-1)}{pq}$ for each $N = pq$).

\subsection{Noise}
\label{ssnoiseres}

\begin{figure}
\begin{center}
\leavevmode
\centerline{\hbox{\psfig{figure=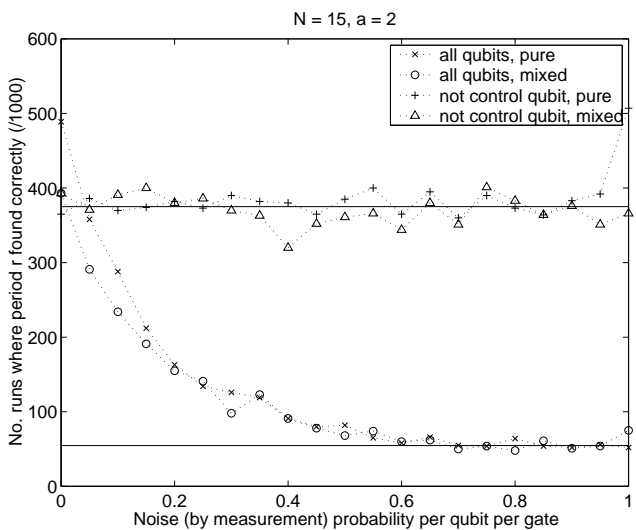,height=7.cm}}}
\end{center}
\caption{The number of runs (out of 1000) of Shor's algorithm for $N=15$ and 
$a=2$ where the period, $r$, is found correctly (as calculated by a continued 
fractions algorithm) when noise, simulated by random measurements, is applied 
independently to each qubit. This number is plotted against the probability that 
a measurement is applied to each qubit after each gate in the algorithm. The 
plot contains both the pure and mixed state versions of Shor's algorithm and for 
noise applied to all of the qubits or all the qubits apart from the control 
qubit. Notice that when noise is not applied to the control qubit the algorithm 
becomes as efficient as the mixed state algorithm (upper solid line). This is 
because noise is not applied during the controlled modulo multiplication gates 
and the period is of the form $r = 2^m$ where $m$ is an integer. The lower solid 
line denotes the efficiency of a completely random algorithm.}
\label{measnoise}
\end{figure}

\begin{figure}
\begin{center}
\leavevmode
\centerline{\hbox{\psfig{figure=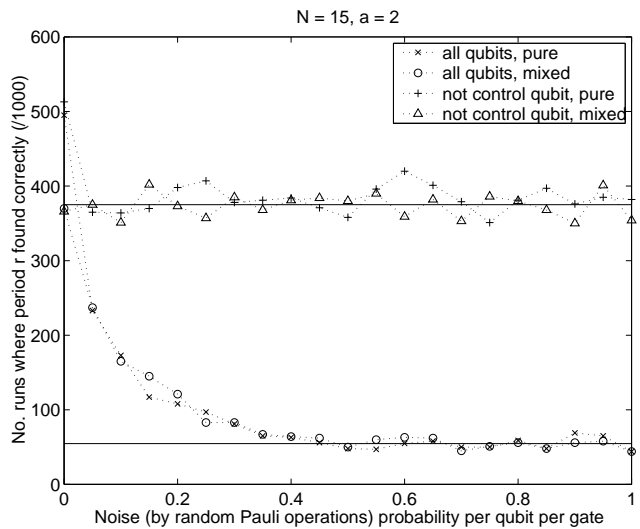,height=7.cm}}}
\end{center}
\caption{As for Fig.\@ \ref{measnoise} but where random Pauli operations are 
applied to each qubit after each gate with the given probability. Again notice 
that when noise is not applied to the control qubit the algorithm becomes as 
efficient as the mixed state algorithm.}
\label{paulinoise}
\end{figure}

\begin{figure}
\begin{center}
\leavevmode
\centerline{\hbox{\psfig{figure=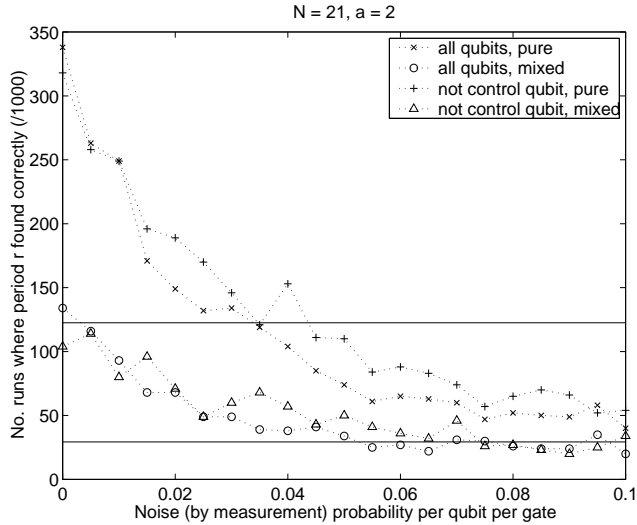,height=7.cm}}}
\end{center}
\caption{As for Fig.\@ \ref{measnoise} but for $N=21$ and $a=2$ where random 
measurement operations are applied to each qubit after each gate with a given 
probability. Here, when noise is not applied to the control qubit the algorithm 
is no longer as efficient as the mixed state algorithm, denoted by the upper 
solid line. The lower solid line shows the efficiency of a completely random 
algorithm.}
\label{n6measnoise}
\end{figure}

\begin{figure}
\begin{center}
\leavevmode
\centerline{\hbox{\psfig{figure=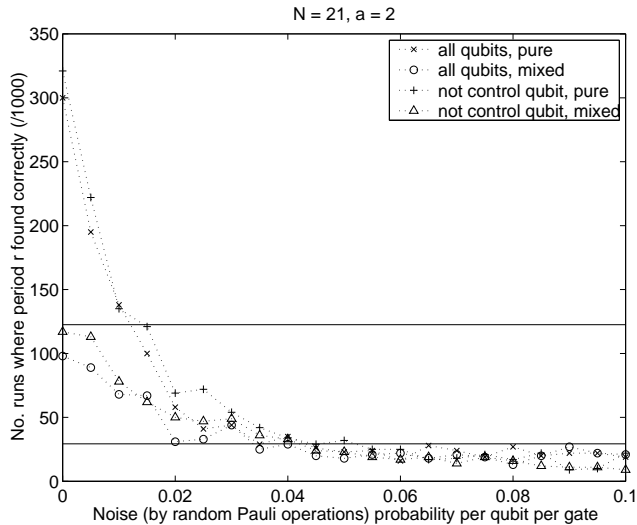,height=7.cm}}}
\end{center}
\caption{As for Fig.\@ \ref{n6measnoise} but where random Pauli operations are 
applied to each qubit after each gate with a given probability. Again notice 
that when noise is not applied to the control qubit the algorithm is no longer 
as efficient as the mixed state algorithm.}
\label{n6paulinoise}
\end{figure}

Next we examine the effect of noise on particular algorithms, namely for $N=15$, 
$a=2$ (Figs. \ref{measnoise} and \ref{paulinoise}) which has period $r=4$ and 
$N=21$, $a=2$ (Figs. \ref{n6measnoise} and \ref{n6paulinoise}) which has period 
$r=6$. Both measurement and Pauli noise were implemented as described in section 
\ref{ssnoise}. Additionally we considered the situation where the noise was not 
allowed to act on the control qubit.

For the $N=15$ case when noise is allowed to act on all qubits we see an 
exponential drop off in the efficiency of the algorithm with increasing noise 
level for both types of noise. However, when noise is not allowed to act on the 
control qubits the efficiency only falls to the efficiency level of the mixed 
state algorithm which, as has been noted, is efficient enough. The is due to the 
way the noise is implemented and the fact that the period is 4. For algorithms 
with period of the form $r=2^m$, for some integer $m$, each of the controlled 
modulo multiplication gates can be applied on any state independently of the 
state of the system up to that point.

The reason is as follows: when the period is of the form $2^m$ the first $n-m$ 
controlled modulo multiplication gates are in fact the identity operation (note, 
however, that individual operations within the gate decomposition of the 
controlled modulo multiplications will not be the identity operation so noise 
would then greatly affect the efficiency of the algorithm) and the measurement 
result after each will be $\ket{0}$. The next gate is now the only relevant 
gate. If the measurement result after this gate is $\ket{1}$ the period will be 
found correctly. In this case
\begin{equation}
\label{bin}
c = \underbrace{??????????}_{m -1 \hbox{ times}} 1 
\underbrace{000\cdots\cdots000}_{n-m \hbox{ times}}.
\end{equation}
where ? = 0 or 1. Now, the fraction $c/t$ will have denominator $r$, independent 
of the results of these $m-1$ remaining measurements.

Noise in the pure state algorithm applied to all but the control qubit, then, 
prepares these qubits in some random pure state (i.e. a mixed state). The first 
$n-m$ stages of the algorithm do nothing but we will correctly find the period 
if the next measurement result is $\ket{1}$, which will occur with the same 
probability as for the mixed state algorithm.

This is not the case for algorithms of other periods of course as the initial 
operations are not identity operations. An example is the $N = 21$, $a = 2$ 
algorithm for which the effects are noise are shown in Figs.\@ \ref{n6measnoise} 
and \ref{n6paulinoise}

These results show us that for general algorithms we cannot increase the mixing 
of algorithms during its running by re-preparing another maximally mixed state 
in the lower register of qubits after each measurement without reducing the 
efficiency of the algorithm considerably - the previous measurements have 
prepared a state which must not be significantly altered before the next stage. 
If the period is of the form $r = 2^m$ we can do this (although having a period 
of this form is presumably exponentially unlikely for increasing $n = \log{N}$) 
but this will have no effect on the entanglement until the last $m$ steps. For 
other algorithms this will reduce the entanglement but will also dramatically 
reduce the efficiency.

\subsection{Mixing of the control qubit}
\label{ssmixcontqres}

\begin{figure}
\begin{center}
\leavevmode
\centerline{\hbox{\psfig{figure=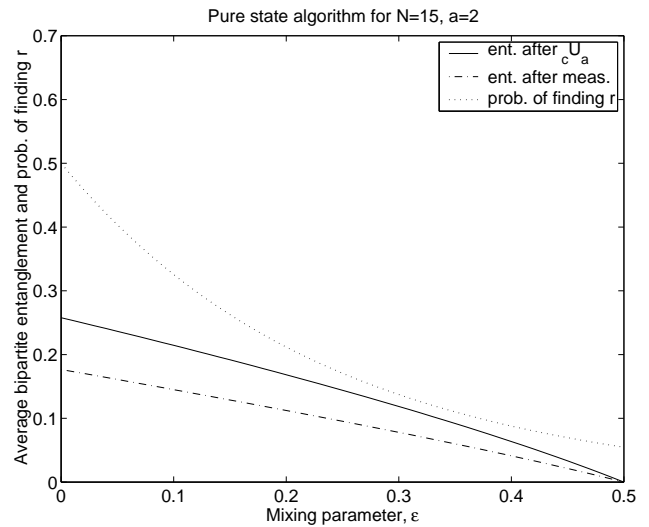,height=7.cm}}}
\end{center}
\caption{Here we show, for the pure state algorithm of $N=15$ and $a=2$, the 
entanglement averaged across all bipartite partitionings and all post -${}_cU_a$ 
and -measurement stages of the algorithm, vs.\@ the mixing parameter, 
$\epsilon$, when the state of the control qubit is repeatedly prepared in the 
mixed state as in section \ref{ssmixing}. Also shown is the probability that the 
algorithm correctly finds the period $r$.}
\label{mixq1pure}
\end{figure}

\begin{figure}
\begin{center}
\leavevmode
\centerline{\hbox{\psfig{figure=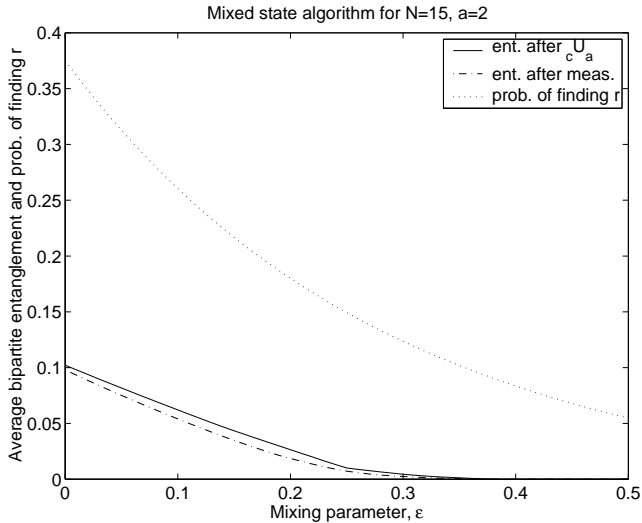,height=7.cm}}}
\end{center}
\caption{As Fig. \ref{mixq1pure} but with the mixed state algorithm of $N=15$ 
and $a=2$. Note that the average entanglement approaches zero before the 
algorithm is maximally mixed, where for the pure state algorithm it is zero only 
when the control qubit is maximally mixed ($\epsilon = 0.5$).}
\label{mixq1mix}
\end{figure}

Let us now look at the effect of mixing the control qubit, as described in 
section \ref{ssmixing}. We will do this for both the 'pure' state algorithm 
(where all other qubits are initially pure) and the mixed state algorithm. 
Figs.\@ \ref{mixq1pure} and \ref{mixq1mix} show the average bipartite 
entanglement throughout the entire algorithm versus the mixing parameter 
$\epsilon$ for the pure and mixed state algorithms with $N=15$ and $a = 2$ 
respectively. Also shown is the probability that each algorithm correctly finds 
the period, $r$.

Notice in particular how the entanglement in the mixed state algorithm 
approaches zero before the algorithm becomes entirely random. For the pure state 
algorithm it does not do so until $\epsilon = 0.5$ is reached. The point at 
which the average entanglement is zero (to machine precision) in the mixed state 
algorithm is around $\epsilon = 0.396$ for the $N=15$, $a=2$ case. This 
illustrates how the randomness in mixed states can completely mask the 
distillable entanglement even before the algorithm becomes entirely random. For 
the pure state case we have mixed two pure state algorithms (with orthogonal 
control qubit states), both of which do produce entanglement, and found that the 
entanglement does not disappear until we mix them maximally. For the mixed state 
algorithm we have also mixed two orthogonal algorithms which both contain 
entanglement but the distillable entanglement has been lost {\em before} we lose 
all information about which algorithm is running.

\begin{figure}
\begin{center}
\leavevmode
\centerline{\hbox{\psfig{figure=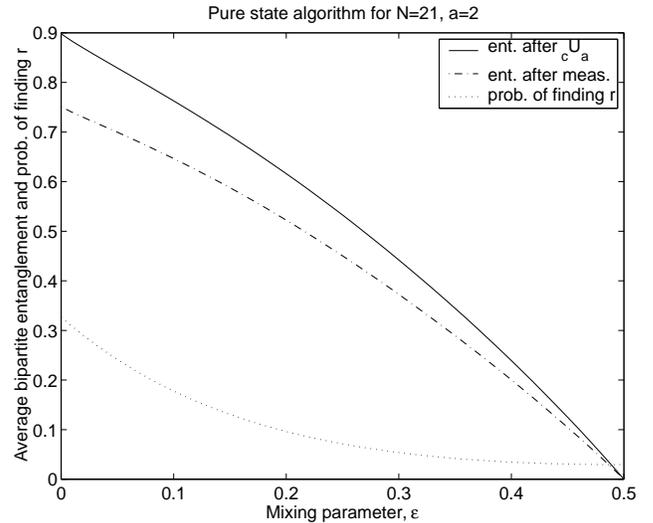,height=7.cm}}}
\end{center}
\caption{For the pure state algorithm of $N=21$ and $a=2$, the entanglement 
averaged across all bipartite partitionings and all post -${}_cU_a$ and 
-measurement stages of the algorithm, vs.\@ the mixing parameter, $\epsilon$, 
when the state of the control qubit is repeatedly prepared in the mixed state as 
in section \ref{ssmixing}. Also shown is the probability that the algorithm 
correctly finds the period $r$.}
\label{mixq1pure2}
\end{figure}

\begin{figure}
\begin{center}
\leavevmode
\centerline{\hbox{\psfig{figure=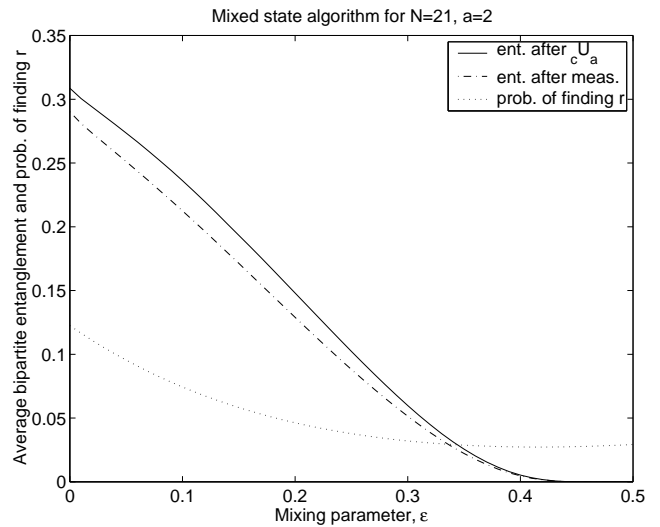,height=7.cm}}}
\end{center}
\caption{As Fig. \ref{mixq1pure2} but with the mixed state algorithm of $N=21$ 
and $a=2$. Again note that the average entanglement approaches zero before the 
algorithm is maximally mixed. The point at which the entanglement is lost occurs 
for a higher value of $\epsilon$ than for the then $N=15$ algorithm.}
\label{mixq1mix2}
\end{figure}

Compare these with similar results for the algorithms with $N=21$ and $a = 2$ in 
Figs. \ref{mixq1pure2} and \ref{mixq1mix2}.

Again we see that the entanglement in the mixed state algorithm is zero before 
the control qubit is maximally mixed, although this point occurs at the higher 
value of around $\epsilon = 0.470$.

\begin{figure}
\begin{center}
\leavevmode
\centerline{\hbox{\psfig{figure=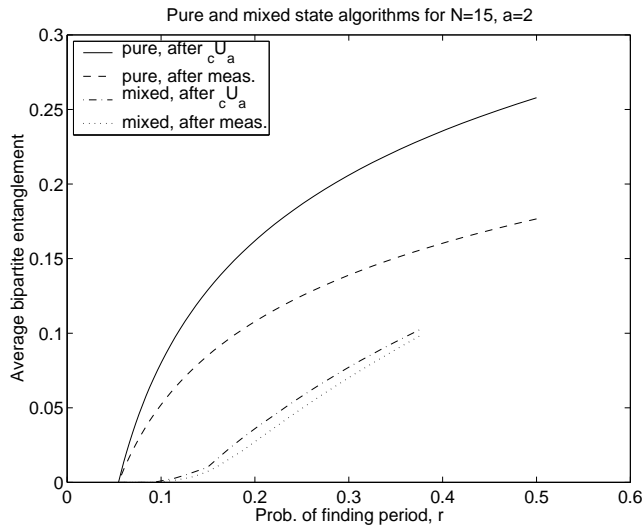,height=7.cm}}}
\end{center}
\caption{Here we show the same set of results as Fig. \ref{mixq1pure} and 
\ref{mixq1mix} but we have plotted the probability of correctly finding the 
period $r$ against the average bipartite entanglement.}
\label{mixq1pvsE}
\end{figure}

\begin{figure}
\begin{center}
\leavevmode
\centerline{\hbox{\psfig{figure=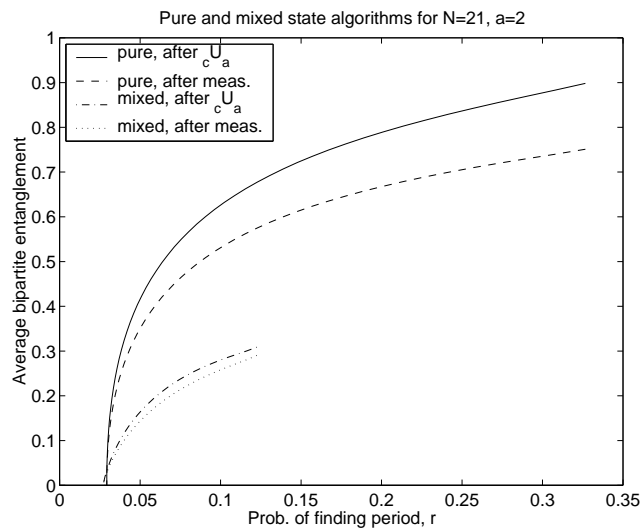,height=7.cm}}}
\end{center}
\caption{Here we show the same set of results as Fig. \ref{mixq1pure2} and 
\ref{mixq1mix2} but we have plotted the probability of correctly finding the 
period $r$ against the average bipartite entanglement.}
\label{mixq1pvsE2}
\end{figure}

Finally we combine the results above to plot the probability of finding $r$ 
against the average entanglement. This is shown in Figs.\@ \ref{mixq1pvsE} and 
\ref{mixq1pvsE2}.

For each particular algorithm, then, we see a definite trend of an increase in 
entanglement giving an increase in the probability of correctly finding the 
period, although initially increasing the entanglement from zero does not 
produce such a large increase in this probability. We also see that the 
entanglement required for a given probability is lower for the mixed state 
algorithm. The maximal probability of correctly finding the period is of course 
lower for the mixed state algorithm but in the limits $p,q \rightarrow \infty$ 
this is negligibly so.

\section{Conclusions}

It has been shown that it is possible to efficiently factorize
with Shor's algorithm using only one initially pure qubit and a
supply of initially maximally mixed qubits. We have also seen that
for algorithms with small numbers of qubits the mixing of the
algorithm remains high throughout. It is then  a natural
question to see whether any entanglement is involved in the
execution of the quantum algorithm. We find that the high degree
of mixing, however, does not preclude the existence of
entanglement and indeed in these example algorithms entanglement
does appear to exist, even when the state of the computer starts
and remains in a highly mixed state.

Conversely, if we try to reduce this entanglement by introducing
further mixing into the control qubit we do reduce the
entanglement in the computer but at the expense of a reduction in
efficiency of the computation. The mixing of the control qubit
also sheds light on the nature of entanglement itself, that is, we
do not need to lose all information about the nature of an
entangled state before we are completely unable to extract
entanglement from it, as we have seen when mixing two orthogonal
entangled mixed state algorithms.

We have also seen that Shor's algorithm operated on mixed states
is nevertheless susceptible to noise of different kinds and the
algorithm does not in general appear to have any increased
robustness to noise (except where the noise model on particular
algorithms is too simplified to be accurate) even though this
algorithm can be run using highly mixed states .

What remains an open question is as to how the above effects
behave for algorithms of increasing numbers of qubits. Presumably
the entanglement does remain for algorithms of large numbers of
qubits and the algorithm remains highly susceptible to noise but
because of the exponential nature of simulating the algorithms
verifying this is an extremely difficult task.

\section*{Acknowledgments}

The authors would like to thank R. Jozsa, S. Virmani, V. Kendon and W. J. Munro for helpful
comments. This work is supported by the United Kingdom Engineering and Physical Sciences
Research Council EPSRC, the Leverhulme Trust, two EU TMR-networks ERB 4061PL95-1412 and ERB
FMRXCT96-0066 and the EU project EQUIP.

\end{multicols}

\end{document}